\documentclass{article}
\usepackage{spconf,amsmath,graphicx,hyperref}

\usepackage{makecell}
\usepackage{comment}
\usepackage{multirow}

\usepackage{caption}

\title{Real-World Adversarial Attacks on RF-Based Drone Detectors}
\name{Omer Gazit$^*$, Yael Itzhakev$^*$, Yuval Elovici, Asaf Shabtai}
\address{Ben-Gurion University of the Negev, Israel\\ {\tt\small \{omergazi,freidiya\}@post.bgu.ac.il, \{elovici,shabtaia\}@bgu.ac.il}}
\begin{document}
%
\maketitle
\def\thefootnote{*}\footnotetext{Equal contribution}\def\thefootnote{\arabic{footnote}}
\begin{abstract}
Radio frequency (RF) based systems are increasingly used to detect drones by analyzing their RF signal patterns, converting them into spectrogram images which are processed by object detection models.
Existing RF attacks against image based models alter digital features, making over-the-air (OTA) implementation difficult due to the challenge of converting digital perturbations to transmittable waveforms that may introduce synchronization errors and interference, and encounter hardware limitations.
We present the first physical attack on RF image based drone detectors, optimizing class-specific universal complex baseband (I/Q) perturbation waveforms that are transmitted alongside legitimate communications.
We evaluated the attack using RF recordings and OTA experiments with four types of drones. Our results show that modest, structured I/Q perturbations are compatible with standard RF chains and reliably reduce target drone detection while preserving detection of legitimate drones.
\end{abstract}

\section{Introduction}
\label{sec:intro}

With unmanned aerial vehicles (UAVs) increasingly used in civilian and defense applications~\cite{basak2021combined}, reliable drone detection near critical infrastructure (e.g., airports, power plants) is essential for the mitigation of risks from unauthorized drone activity~\cite{basak2021combined,bisio2021localization}
Radio frequency (RF) based detection, which can leverage drones’ spectral–temporal signatures, has shown robustness under non-line-of-sight conditions~\cite{elyousseph2024robustness,bisio2021localization}.
Converting complex baseband (I/Q) signals to time frequency images (i.e., spectrograms) via Short-Time Fourier Transform (STFT)~\cite{basak2021combined} enables the use of computer vision object detection (OD) frameworks like YOLO~\cite{redmon2016you} and Faster R-CNN~\cite{ren2015faster} for drone identification and localization in multi-emitter environments~\cite{basak2021combined,zhao2024drone,zhao2023anchor}.
However, OD frameworks are vulnerable to adversarial attacks where small perturbations can impede detection and induce false negatives~\cite{amirkhani2023survey}.

Despite the growing use of image based OD models for congested RF scenarios, prior RF adversarial studies focused on targeting classification models, and most of them optimized perturbations on digital spectrogram pixels or learned embeddings~\cite{kokalj2019targeted,li2024slpa,papangelo2024adversarial}.
However, converting these digital modifications back to transmittable I/Q waveforms is challenging due to bandwidth limitations, power constraints, and hardware limitations.
In addition, validation has been predominantly digital in these studies, while real-world transmission introduces complexities such as carrier frequency offsets, multipath propagation, and timing misalignment that could diminish attack effectiveness~\cite{kokalj2019adversarial}, leaving RF detection systems' vulnerability to adversarial attacks not fully understood.

We introduce the first physical adversarial attack targeting drone transmission specifically designed for RF based OD models.
Rather than optimizing perturbations in the digital feature space, our approach generates adversarial waveforms directly in the I/Q domain and emits them over-the-air (OTA) from an independent transmitter.
This creates interference that is combined with legitimate drone signals but eliminates the need for the complex digital-to-RF conversion and ensures compatibility with standard RF hardware.
We perform both digital and physical evaluation using the same optimized I/Q-domain CUAP across various OD architectures, demonstrating the generalizability of I/Q domain adversarial perturbation in RF detection systems.

We summarize the contributions of our work as follows:
(1) To the best of our knowledge, this is the first work to demonstrate a physical adversarial attack specifically targeting image based OD models, addressing the unique challenges of multi-emitter detection scenarios.
(2) We introduce a novel method that optimizes perturbations directly in the I/Q domain, ensuring physical realizability and compatibility with standard RF transmission chains.
(3) We conduct comprehensive OTA experiments using RF transmissions from four drone platforms, validating that modest, structured I/Q perturbations learned digitally reliably degrade detection performance while overcoming real-world wireless challenges such as timing misalignment and channel effects.

\section{Related Work }
\label{sec:related_work}

\noindent\textbf{Adversarial ML attacks on RF based models.}
Adversarial attacks on RF based deep learning systems can be categorized according to several key properties: 
\textit{(i) input dependency:} per-frame perturbations optimized for each frame versus universal adversarial perturbations (UAPs) that optimize one perturbation for all frames;
\textit{(ii) temporal alignment:} synchronous attacks requiring physical clock alignment with target transmitters and frame-aligned perturbations versus asynchronous attacks with time-shift invariant perturbations that work regardless of signal timing within frames;
\textit{(iii) model task:} classification versus OD tasks, applied to signal modulation (SM), fingerprint identification (FP-ID), or protocol classification (PC).

All prior RF adversarial research targeted classification models rather than OD frameworks.
Most attacks focus on per-sample perturbations for SM classification tasks~\cite{kokalj2019targeted,flowers2019evaluating,chew2022adversarial}, requiring individual perturbation generation for each input, reducing attack efficiency.
UAPs are more efficient~\cite{sadeghi2018adversarial,kim2021channel} but have been limited to classification tasks and typically target all signal classes indiscriminately. Class-specific UAPs (CUAPs)~\cite{sandler2022real} provide more targeted attacks but have not been extended to RF FP-ID or OD scenarios.
Recent work on FP-ID includes per-sample methods~\cite{liu2023robust} and digital image/feature space attacks~\cite{papangelo2024adversarial,li2024slpa} that lack OTA validation.

\noindent\textbf{Physical layer validation and practical deployment.}
Only a few studies conducted physical experiments, performing a per-sample attack~\cite{kokalj2019targeted} and UAP on an SM classifier~\cite{sandler2022real,shtaiwi2023implementation}, leaving OTA validation for RF FP-ID tasks, especially on OD models, unexplored. Additionally, most existing attacks assume perfect knowledge of target architectures, with limited exploration of transferability~\cite{kim2021channel,chew2022adversarial}.
Critically, CUAPs have not been studied in realistic, asynchronous multi-emitter conditions where selective targeting of specific drone classes must preserve detection of legitimate traffic.

\section{The Proposed Attack Framework}
\label{sec:method}

We present a method for optimizing a \emph{CUAP} (an I/Q waveform) to suppress detections of a chosen target class in RF image based OD models while preserving the detection of non-target classes. The adversarial waveform is designed to be: (i) \emph{input-independent} and reusable across different inputs and channel conditions; (ii) \emph{time-shift invariant} through random circular shifts and tiled parameterization; and (iii) \emph{OTA-realizable} by optimizing directly in the I/Q domain under practical power constraints and asynchronous conditions.

\subsection{Adversarial Waveform Design}

\label{sec:process}
Let $x_{clean}$ be the recorded I/Q signal of the drone, $\delta$ be the learned adversarial I/Q waveform, and $x_{adv}$ be the combined I/Q signal received by the detector. The adversarial signal is defined as:
\begin{equation}
x_{adv} = x_{clean} + \delta
\label{eq:def}
\end{equation}

\begin{figure}
    \centering
    \includegraphics[width=\columnwidth]{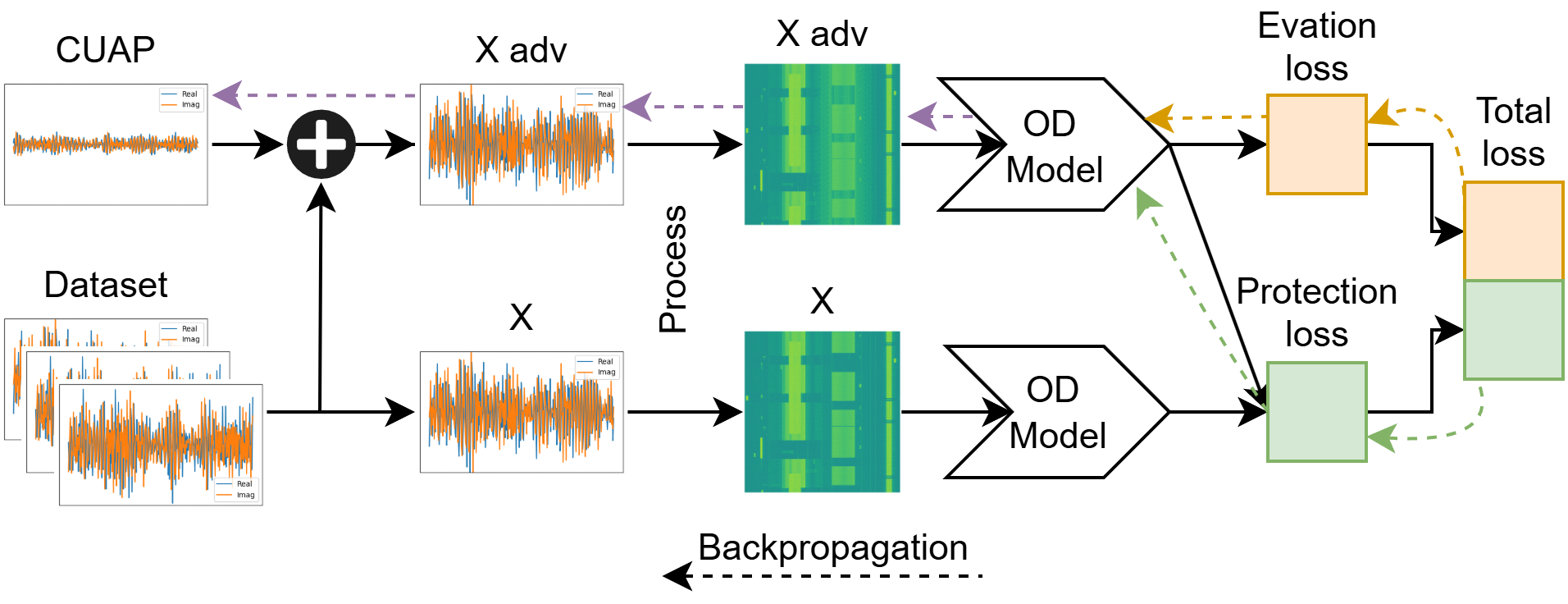}
    \caption{\small Overview of our method’s pipeline.}
    \label{fig:overview}
\end{figure}

\noindent\textbf{Signal processing pipeline.}
Both clean and adversarial signals undergo identical processing to reach the detector: each signal is divided into segments of 1,024 samples using Hann windowing. An STFT is applied with 1,024-point windows and hop length $h=1,024$ (no overlap), yielding $1,024\times1,024$ complex time frequency coefficients $S$. The magnitude $|S|$ is converted to decibels $M=10\log_{10}(|S|+\varepsilon)$, clipped to a constant global range $[M_{\min},M_{\max}]$ determined across all recordings in the training set for consistent normalization, and linearly mapped to $[0,1]$, resulting in the final spectrogram input to the detector. The adversarial waveform is generated as an I/Q stream which is then being added to the recorded I/Q stream and processed together into one spectrogram (as shown in Fig.~\ref{fig:overview}). This pipeline is fully differentiable, enabling gradient flow from detector $\mathcal{D}$ back to the I/Q domain for end-to-end optimization of perturbation $\delta$.

\noindent\textbf{Perturbation constraints and design.}
To ensure a practical adversarial attack, we bound the perturbation power by enforcing a minimum signal perturbation ratio (SPR) of 10\,dB, defined as the average clean signal power to average perturbation power ratio.
For robustness and reduced synchronization requirements, we learn a short I/Q perturbation $\delta$ of size $64{\times}1,024$ samples (64 STFT frames, 1,024 samples per frame). We tile this perturbation along time with random circular shifts applied in the I/Q domain, where the shift offset is uniformly sampled from $[0, |\delta|)$. Each $1,024{\times}1,024$ input spectrogram is thus covered by a randomly shifted version of the 64-frame tile rather than frame-aligned repetitions. Physically, the base waveform is transmitted OTA in a continuous loop.
OTA timing uncertainty is modeled during training via an intra-hop random circular shift (described below).

\subsection{Optimization Objective}
Perturbation $\delta$ is optimized digitally through an iterative procedure using standard gradient descent, and then reused unchanged for both digital and OTA evaluations, aiming to achieve our main objective of reducing target class confidence while preserving non-target classes as follows:

\noindent\textbf{Suppressing the detection of the target class.}
To selectively suppress the target class $c^*$ and avoid attacking false positives, we retain only detections that spatially align with the target class ground truth:
Let $\mathcal{I}=\{1,\dots,N\}$ index all pre-NMS predictions of the detector, where each prediction $(b_i, s_{i,1}, \ldots, s_{i,C})$ contains the bounding box $b_i$ and the confidence score $s_{i,c}$ for each class $c\in C$, 
and let $\mathcal{G}_{c^*}$ be the ground-truth boxes of the target class $c^*$.
We define the set of predictions that match the ground truth using intersection-over-union (IoU) threshold $\tau = 0.5$:
\begin{equation}
\mathcal{S}_{c^*}
\;=\;
\big\{\, i\in\mathcal{I} \;:\; \exists\, g\in\mathcal{G}_{c^*}\ \text{s.t.}\ \mathrm{IoU}(b_i,g)\ge\tau \,\big\}
\end{equation}
The \textit{evasion loss} drives the target class confidence toward zero and minimizes the probability of detection:
\begin{equation}
\mathcal{L}_{\text{evade}} = 
\frac{1}{|\mathcal{S}_{c^*}|} \sum_{i \in \mathcal{S}_{c^*}} -\log\bigl(1 - s_{i,c^{*}}\bigr)
\end{equation}

\noindent\textbf{Maintaining the detection of untargeted classes.}
We include another component to maintain consistent detection of non-target classes by minimizing the output divergence:
\begin{equation}
\mathcal{L}_{\text{protect}} = \frac{1}{N(C-1)} \sum_{i=1}^{N} \sum_{\substack{c=1 \\ c \neq c^{*}}}^{C} \big\| s_{i,c}(x_{\text{adv}}) - s_{i,c}(x_{\text{clean}}) \big\|
\end{equation}
The final objective combines both terms:
\begin{equation}
\mathcal{L}(x_{adv}) = \mathcal{L}_{\text{evade}}(x_{adv}) + \lambda \cdot \mathcal{L}_{\text{protect}}(x_{adv})
\end{equation}
with $\lambda=2$ controlling the trade-off between attack effectiveness and non-target preservation.

In each training iteration, we (i) uniformly sample a time offset over the $64{\times}1,024$ perturbation tile in the I/Q domain and apply a circular shift to $\delta$ prior to mixing with $x_{\text{clean}}$, emulating timing asynchronization;
(ii) compute the losses and update the parameters; and 
(iii) normalize $\delta$ to satisfy $\mathrm{SPR}\le 10\,\mathrm{dB}$.

\vspace{-3pt}
\section{Evaluation Setup}
\label{sec:eval}
\vspace{-3pt}

\subsection{Dataset}
The evaluation was performed on a dataset of drones' RF recordings (I/Q stream) that we collected.

\noindent\textbf{Training data collection.}
We collected RF baseband I/Q data using an \textit{Analog Devices ADRV9009} transceiver interfaced to a \textit{Xilinx ZCU102} evaluation board. All training recordings were obtained in a shielded room (Faraday cage) to suppress external interference and ensure signal integrity. The receiver was configured with a local oscillator at 2.45\,GHz, a complex sampling rate of 122.88\,$\mathrm{MS/s}$, and an effective analog bandwidth of 100\,MHz, fully covering the 2.4\,GHz ISM band used by our drones.
Four types of drones (Mavic~2~Zoom, Mavic~Pro, Mavic~Air, and Matrice~600) were operated in idle-on mode separately on multiple RF channels; each drone’s transmissions were recorded, converted to spectrograms using the pipeline described in Sec.~\ref{sec:process}, and segmented into detector input frames of size $F\times T = 1,024\times1,024$ (frequency $\times$ time).

For each class of drones we augmented the recordings by digitally mixing baseband I/Q signals from the recording sessions of different drones, and applying random carrier frequency offsets to shift transmissions across the 2.4\,GHz band.
To increase training data diversity, additive white Gaussian noise (AWGN) was injected at controlled SNRs.
To match the three-channel input of image based OD models, we constructed a pseudo-RGB tensor by duplicating the normalized single-channel spectrogram along the channel dimension. Formally, if $X\in[0,1]^{F\times T}$ denotes the magnitude (dB) spectrogram after normalization, we feed $\tilde{X}=\mathrm{stack}(X,X,X)\in[0,1]^{3\times F\times T}$. No color mapping or additional preprocessing was applied.
The final dataset divided into target, surrogate, and test subsets in proportions of 40\%, 40\%, and 20\%, respectively.

\noindent\textbf{Training data annotation.}
While recording, we added the occupied band range reported by the drone to the metadata.
Then we isolated those frequencies and applied an energy detector to extract time annotations of the transmission bursts and manually validated the bounding boxes.

\noindent\textbf{Attack set generation.}
We assume an adversary that aims to prevent their drone's detection by OD systems.
They can control their drone’s transmit channel (i.e., center frequency) but have no control of other drones' transmission channels.
Accordingly, we recorded the target drone OTA in a realistic, noisy environment; other drones' signals were digitally mixed into the I/Q stream at random channels across the 2.4\,GHz ISM band.
This setup reflects a realistic threat model in which the adversary transmits
an additional OTA adversarial signal but cannot influence coexisting signals.
The attack set used to train the adversarial waveform was split into 70\%, 15\%, and 15\% for the training, validation, and test subsets, respectively.

\vspace{-5pt}
\subsection{Attacked Detection Models}
\vspace{-3pt}
The detectors were trained on magnitude (dB) spectrograms derived from the drones' recordings, using the pipeline in Sec.~\ref{sec:process}.
We evaluate our attack on five pretrained object detectors (YOLOv5, v8, v9, v11~\cite{redmon2016you}, and Faster R-CNN~\cite{ren2015faster}) fine-tuned on the annotated spectrograms from the \textit{target} subset.
The detectors' performance on the clean \textit{target} subset is reported in the \textit{Clean} row of Table~\ref{table:digital}.
To assess the transferability of adversarial perturbations across architectures and data distributions, surrogate detectors were fine-tuned on the \textit{surrogate} subset and used to train the perturbations.

\vspace{-5pt}
\subsection{Evaluation Metrics}
\vspace{-3pt}
For digital evaluations, we report average precision (AP) for the target class and mean AP (mAP) across all non-target classes, where AP represents the area under the precision–recall curve at an IoU threshold of 0.5.
For physical experiments, where precise frame-level annotations are unavailable, we quantify performance using the missed detection rate (MDR) for the class, reflecting the same attack objective as AP, where non-target classes are aggregated in a single class:
\begin{equation}
\mathrm{MDR}(\text{class})=\frac{\#\,\text{missed signals (class)}}{\#\,\text{total signals (class)}}
\label{eq:mdr}
\end{equation}
We use a fixed confidence threshold of 0.4 in which the produced mAP $=0.947$ on clean data across detectors.
A signal is considered \emph{missed} if, after applying a threshold of $0.4$ and applying NMS, no prediction achieves an $\mathrm{IoU}\ge 0.5$.
A successful attack is thus reflected by a high MDR on the target class and a low MDR on non-target classes (i.e., suppress target detection and preserve non-target detection).

\vspace{-3.5pt}
\subsection{Experimental Results}
\vspace{-1pt}
We focus on the the results for the Mavic~2~Zoom due to space limitations; consistent patterns were observed for the other evaluated drone platforms.

\noindent\textbf{Digital attacks.}
For the digital evaluation, we digitally injected the CUAP into the I/Q stream prior to spectrogram formation (Eq.~\ref{eq:def}).
As baselines we report the results obtained on \emph{clean} I/Q signals and with random \emph{AWGN}.
We considered the following four attacker-knowledge scenarios: 
\emph{White-box (WB):} per-architecture CUAP with full access to the target detector’s weights;
\emph{Gray-box (GB):} CUAP optimized on a surrogate with the same architecture using a disjoint training subset;
\emph{Closed-set (CS):} CUAP optimized jointly on all surrogates;
\emph{Black-box (BB):} leave-one-architecture-out training where the held-out architecture serves as the BB target.

As can be seen in Table~\ref{table:digital}, for the Mavic~2~Zoom in the \textit{WB} setting, the target class's AP consistently collapsed to near zero while preserving the non-target classes' mAP, indicating selective suppression rather than global degradation.
In the \textit{GB} and \textit{CS} settings, transferability varies across models, with the \textit{CS} scenario generally outperforming the \textit{GB}.
The \textit{BB} setting produced the largest residual target AP value yet still resulted in substantial suppression relative to that of the clean/AWGN settings.
AWGN has a negligible effect on the target class, underscoring that the observed failures are adversarial rather than power-induced.
For other drone platforms, similar results were typically obtained in the \textit{CS} setting.
\begin{table}[t!]
\centering
\small
\setlength{\tabcolsep}{4pt} 
\renewcommand{\arraystretch}{1.1}
\resizebox{0.93\columnwidth}{!}{%
\begin{tabular}{|c|c||c|c|c|c|c| }
 \hline
 & & \multicolumn{5}{|c|}{\textbf{Model Architecture}} \\
 & & \multicolumn{5}{|c|}{Target class AP $\downarrow$ / Non-target classes mAP $\uparrow$} \\
 \hline
 Drone & Setting & YOLOv5 & YOLOv8 & YOLOv9 & YOLOv11 & FRCNN\\
 \hline
 \multirow{6}{*}{Mavic~2~Zoom} & Clean      & 0.95 / 0.98 & 0.98 / 0.99 & 0.92 / 0.98 & 0.97 / 0.99 & 0.9 / 0.96 \\
 & AWGN     & 0.96 / 0.98 & 0.98 / 0.99 & 0.93 / 0.98 & 0.96 / 0.99 & 0.9 / 0.97 \\
 & CUAP (WB)  & 0.08 / 0.86 & 0.0 / 0.81  & 0.0 / 0.96  & 0.0 / 0.91  & 0.07 / 0.86 \\
 & CUAP (GB)  & 0.4  / 0.94 & 0.57 / 0.95 & 0.51 / 0.94 & 0.62 / 0.96 & 0.35 / 0.92 \\
 & CUAP (CS)  & 0.34 / 0.94 & 0.54 / 0.97 & 0.26 / 0.93 & 0.38 / 0.96 & 0.23 / 0.92 \\
 & CUAP (BB)  & 0.48 / 0.94 & 0.64 / 0.95 & 0.33 / 0.94 & 0.6  / 0.98 & 0.53 / 0.96 \\
 \hline \hline
 Matrice~600 & CUAP (CS)  & 0.32  / 0.84 & 0.8 / 0.91 & 0.49 / 0.82 & 0.03 / 0.83 & 0.42 / 0.97 \\
 \hline
 Mavic~Air & CUAP (CS)  & 0.04 / 0.96 & 0.33 / 0.96 & 0.25 / 0.94 & 0.22 / 0.97 & 0.8 / 0.97 \\
 \hline
  Mavic~Pro & CUAP (CS)  & 0.9 / 0.84 & 0.55 / 0.94 & 0.51 / 0.82 & 0.23  / 0.85 & 0.92 / 0.96 \\
 \hline
\end{tabular}}
\caption{\small AP under \textbf{digital} attacks (for SPR = 10~dB) in the following settings: clean, AWGN, WB, GB, CS, and BB.}
\label{table:digital}
\end{table}

\begin{figure}
    \captionsetup{skip=4pt}
    \centering
    \includegraphics[width=0.75\columnwidth, height=0.075\textheight]{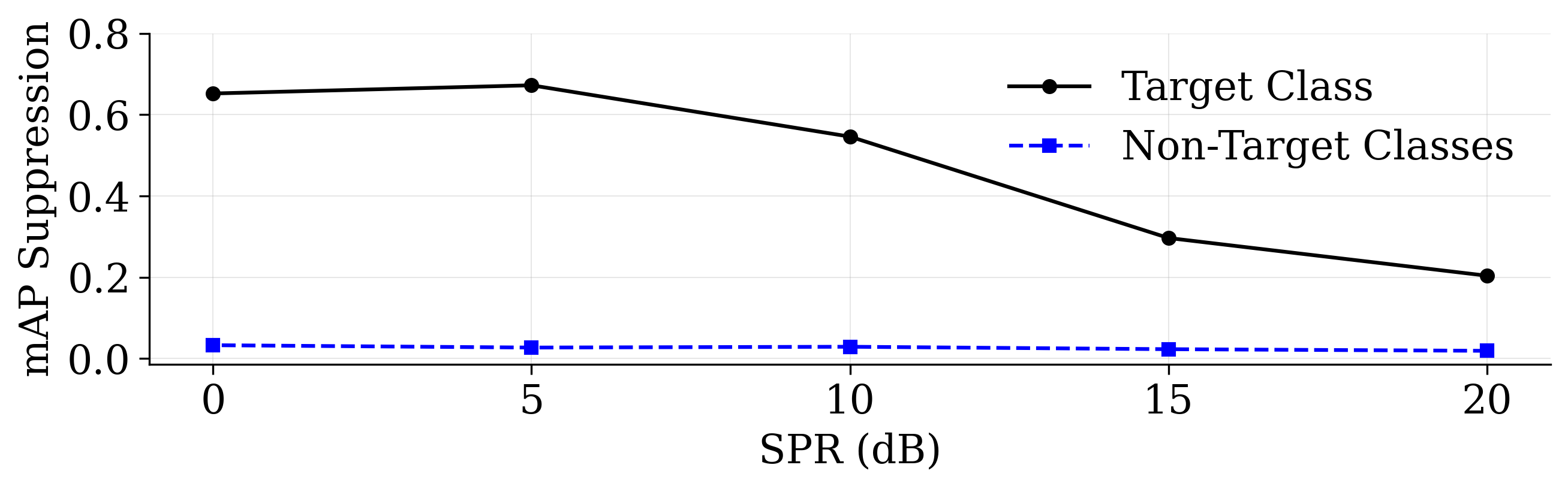}
    \caption{\small The mAP as a function of the SPR power (dB).}
    \label{fig:spr}
\end{figure}

\noindent\textbf{SPR analysis.}
Fig.~\ref{fig:spr} presents the attack effectiveness as a function of the SPR.
As can be seen, as the SPR decreases, the target class's AP decreases monotonically while the non-target classes' mAP remains unchanged, confirming that our class-specific approach improves evasion without degrading non-target class detection. The 10~dB SPR constraint balances effective drone suppression with low detectability.

\noindent\textbf{Physical attacks.}
Due to space limitations, we present only the \emph{closed-set} CUAP, which exhibited the strongest cross-architecture transfer in digital evaluations to demonstrate OTA feasibility in a realistic multi-emitter environment.
Following the described threat model, the target drone operated in the same predetermined channel, while the other three drones independently selected transmission channels, eliminating spectral hopping and isolating OTA effects. The optimized $\delta$ was transmitted alongside legitimate communications, with transmit power adjusted to match the received perturbation levels from digital experiments.
Drones were placed at fixed distances (3\,m, 5\,m, and 7\,m) from the receiving antenna to assess range effects.
\emph{Clean} and \emph{Attacked} recordings were acquired consecutively under identical conditions due to non-repeatable OTA timing and phase.

The results presented in Table~\ref{table:physical} highlight the transferability of our digital optimization approach to a physical attack designed for the Mavic~2~Zoom.
The target class MDR reached near-complete suppression ($\ge 99\%$) on most models across distances, with YOLOv9 at 3\,m and FRCNN as exceptions.
Detection of the non-target classes remained largely preserved (MDR degradation $\le 18.3\%$), confirming a class-specific targeting effect.
These OTA results demonstrate that I/Q domain perturbations retain adversarial effects after propagation through physical RF channels across varying ranges.

\begin{table}[t!]
\centering
\small
\setlength{\tabcolsep}{4pt} 
\renewcommand{\arraystretch}{1.1}
\resizebox{0.76\columnwidth}{!}{%
\begin{tabular}{|c|c||c|c|c|c|c| }
 \hline
 & \multicolumn{6}{|c|}{\textbf{Model Architecture}} \\
 & \multicolumn{6}{|c|}{Target class MDR \% $\uparrow$ / Non-target classes MDR \% $\downarrow$} \\
 \hline
 Distance & Setting & YOLOv5 & YOLOv8 & YOLOv9 & YOLOv11 & FRCNN\\
 \hline
\multirow{2}{*}{3m} &Clean     & 1.7 / 3.6  & 1.0 / 2.8 & 0.7 / 2.3 & 0.4 / 5.2 & 3.9 / 14.7 \\
 &Attacked  & 100 / 6.8 & 100 / 2.9 & 71.5 / 3.4 & 100 / 5.1 & 90.3 / 6.9 \\
 \hline
 \multirow{2}{*}{5m} &Clean     & 0.9 / 3.3  & 0.2 / 3.6 & 5.2 / 1.7 & 1.7 / 3.9 & 2.9 / 9.4 \\
 &Attacked  & 100 / 6.5 & 100 / 5.8 & 100 / 12.3 & 100 / 16.0 & 90.7 / 3.7 \\
 \hline
 \multirow{2}{*}{7m} &Clean     & 3.0 / 2.1  & 1.3 / 11.0 & 9.6 / 1.1 & 3.0 / 12.3 & 7.6 / 11.0 \\
 &Attacked  & 100 / 6.7 & 99.5 / 13.6 & 100 / 5.7 & 100 / 18.3 & 87.5 / 3.3 \\
 \hline
\end{tabular}}
\captionsetup{skip=9pt}
\caption{\small\textbf{OTA} performance (CS setting, Mavic~2~Zoom) at varying distances.
High MDR for the target class indicates successful attack.}
\label{table:physical}
\end{table}

\vspace{-3pt}
\section{Discussion and Conclusion\label{sec:conclusion}}

\vspace{-4pt}
This work presents the first physical adversarial attack against image-based RF drone detectors by learning CUAPs in the I/Q domain, enabling OTA implementation.
Our evaluation demonstrates selective suppression of the target class with minimal impact on non-target classes and strong cross-architecture transfer.
These proof-of-concept results, limited to an office environment, idle-on drones, and fixed distances, reveal fundamental vulnerabilities in RF-based detection systems and motivate robust defenses such as adversarial training and RF anomaly detection.


\vfill\pagebreak

\bibliographystyle{IEEEbib}
\bibliography{mybib}

\end{document}